\documentstyle[aps,prl,multicol,epsf]{revtex}
\renewcommand{\narrowtext}{\begin{multicols}{2} \global\columnwidth20.5pc}
\renewcommand{\widetext}{\end{multicols} \global\columnwidth42.5pc}
\renewcommand{\v}[1]{{\bf #1}}

\newcommand{\sgn}{{\rm sgn}}

\newcommand{\ba}{\begin{eqnarray}}
\newcommand{\ea}{\end{eqnarray}}
\newcommand{\be}{\begin{equation}}
\newcommand{\ee}{\end{equation}}
\newcommand{\nn}{\nonumber\\}

\newcommand{\cL}{ {\cal L}}
\newcommand{\cG}{{\cal G}}

\newcommand{\bscco}{Bi$_2$Sr$_2$CaCu$_2$O$_{8+\delta}$}
\begin{document}
\draft
\title{
Signature of Spin Collective Mode in Local Tunneling Spectra \\
of a d-wave Superconductor}
\author{Jung Hoon Han}
\address{Department of Physics, Konkuk University\\
1 Hwayang-dong, Kwangjin-gu, Seoul 143-701, Korea}

\maketitle \draft
\begin{abstract}
We consider the influence of magnetic excitations on the local
density of states in the $d$-wave superconductor. The magnetic
susceptibility is calculated within the renormalized $t-t'-J$
model and its influence on the quasiparticle self-energy is
considered using a minimal model originally proposed by
Polkovnikov {\it et al.}[cond-mat/0203176]. We find the local
density of states possess periodic components both along $(\pi,0)$
and $(\pi,\pi)$ directions with the associated wavevectors
changing in magnitude as the quasiparticle energy is varied.
Comparison with the STM experiment reveals that the calculated
LDOS modulation is inconsistent with the measured data.
\end{abstract}

\narrowtext {\it Introduction}: Observation of discernible
``checkerboard'' patterns in the local density of states (LDOS)
has been reported for a superconducting \bscco~ (BSCCO) compound
in the mixed state, with the spatial periodicity close to four
lattice constants \cite{hoffman1}. Howald {\it et al.} reported
the same periodicity exists for the LDOS in the same compound but
without the magnetic field\cite{howald}. Hoffmann {\it et al.} in
turn argued that this periodicity is in fact a function of energy,
{\it i.e.} as the probing bias of the scanning tunnelling
microscope (STM) is varied the spatial modulation periodicity also
changes systematically\cite{hoffman2}.

A number of theories have emerged in response to such striking set
of observations\cite{neworder,wl,pvs}. Some of these theories
emphasize the possible emergence of a new order parameter near the
vortex core\cite{neworder} but, in view of the latest experimental
finding\cite{hoffman2} may have difficulty to explain how an
induced order parameter can exhibit a length scale that depends
smoothly on energy. An alternative explanation advanced by Wang
and Lee (WL) ascribes the periodicity to the quasiparticle
scattering due to localized impurities\cite{wl}. In this approach
scattering processes connect parts of the underlying Fermi
contours at different wavevectors as the quasiparticle energy is
varied.

Quite distinct from the proposal of WL is the theory of
Polkovnikov, Vojta, and Sachdev(PVS)\cite{pvs} who showed in a
simple model that a dynamic spin fluctuation mode with a distinct
ordering wavevector $\v Q$ may lead to observable periodic
modulations in the local STM spectra at twice $\v Q$. They also
invoke the local impurity pinning, which is necessary to break the
translational symmetry and give rise to spatially varying LDOS. In
principle the mechanism proposed by PVS does not depend on the
presence of an external magnetic field.

Dynamical spin fluctuations are observable in a number of families
of cuprates. It has been shown that magnetic absorption peaks
occur at incommensurate wavevectors $(\pi,\pi\pm\delta)$, and
$(\pi\pm\delta,\pi)$ with $\delta$ varying with the absorption
energy\cite{neutron}. Brinckmann and Lee (BL)\cite{bl} argued that
the underlying Fermi surface topology in the superconducting state
is responsible for such incommensurate response. Within the
Gutzwiller-renormalized $t-t'-J$ model they indeed found the
absorption peaks at $(\pi\pm\delta,\pi)$, and $(\pi,\pi\pm\delta)$
while at even lower energies, the incommensurate peaks appear in
the diagonal direction, $(\pi\pm \delta',\pi\pm\delta')$.

Coupled with the idea of PVS, it is worthwhile to ask whether such
incommensurate magnetic spectrum will also modulate the LDOS at
wavevectors that vary with the energy. In this paper, we examine
this issue using the magnetic spectrum calculated from the
renormalized $t-t'-J$ model. Assuming a simple interaction scheme
between the spin of the quasiparticles and the magnetic excitation
we evaluate the quasiparticle self-energy and thence the spatially
varying LDOS. Ultimately our calculation is aimed at a comparison
with the latest experimental data\cite{hoffman2}, and in the
process hope to figure out to what extent, if at all, the magnetic
scenario for the observed periodicity is viable.
\\

{\it Methods}: We adopt the renormalized $t-t'-J$ model\cite{bl}
as the starting point: \ba
&&H=-t_{eff}\sum_{ij}f^\dag_{j\sigma}f_{i\sigma}+
t'_{eff}\sum_{ij}f^\dag_{j\sigma}f_{i\sigma}-\mu \sum_i
f^\dag_{i\sigma}f_{i\sigma}\nn &&~~~-{J\over4}\sum_{ij}
(\Delta^*_{ij}\epsilon_{\alpha\beta}f_{i\alpha}f_{j\beta}\!+\!h.c.)+
{J_{eff}\over2}\sum_{ij}S_i \cdot S_j. \label{effective-H} \ea In
the above, $ t_{eff}=tx+(J/4)\sum_\sigma\langle
f^\dag_{j\sigma}f_{i\sigma}\rangle,  t'_{eff}=t' x,
\Delta_{ij}=\langle\epsilon_{\alpha\beta}f_{i\alpha}f_{j\beta}\rangle$,
and $x$ is the average doping. $J_{eff}$ is the residual spin-spin
interaction between quasi-particles.

\begin{figure}
\centering \hskip -0.0cm
\epsfxsize=7cm\epsfysize=4cm\epsfbox{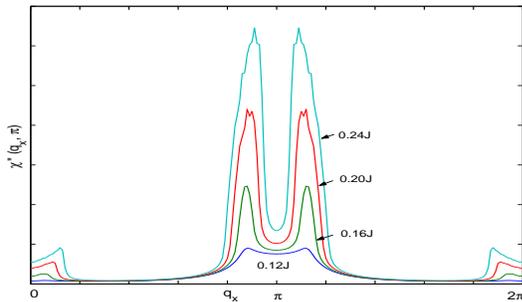}
\caption{Imaginary part of the magnetic susceptibility, $\chi''
(\v q, \nu)$ for $q_y = \pi$, and $0<q_x <2\pi$ calculated from
the renormalized $t-t'-J$ model with $J_{eff}=0.34J$. Different
plots correspond to $\nu=0.12J, 0.16J, 0.20J$, and $0.24J$.}
\label{chi-k}
\end{figure}

From the above mean-field Hamiltonian one calculates the
zero-temperature magnetic susceptibility $\chi_0 (\v q,i\nu)$, and
through the residual spin-spin interaction $J_{eff} S_i \cdot
S_j$, the renormalized spin susceptibility \be \chi (\v q,
i\nu)={\chi_0 (\v q, i\nu) \over 1+J(\v q)\chi_0 (\v q,
i\nu)},\label{chi-RPA}\ee for $J(\v q) =J_{eff}(\cos q_x \!+\!\cos
q_y )$. In PVS's treatment the magnetic susceptibility is
approximated by a phenomenological form, $\chi_{PVS}(\v
q,i\nu)^{-1}\propto \nu^2+\Delta_s^2 + c^2 (\v q-\v Q)^2$, where
$\Delta_s$ is the spin gap, and $\v Q$ is the ordering wavevector.
On the other hand, the $t-t'-J$ model result for the renormalized
magnetic susceptibility is known to exhibit a continuous set of
ordering wavevectors with respect to energy\cite{bl}. We show a
horizontal scan of the imaginary part $\chi''(\v q,\nu)$ with $q_y
=\pi$ for a series of energies $\omega$ in Fig. \ref{chi-k}. The
incommensurate peaks near $q_x =\pi$ are obvious, and both these
peaks and the less pronounced peaks near $q_x =0$ move toward $q_x
=\pi$ when the energy is raised. In particular the peaks at
$(\pi\pm\delta(E), \pi)$ may give rise to a modulation in the
charge sector with twice the wavevector\cite{pvs,GL}, {\it i.e.}
at $(2\pi\pm2\delta(E),2\pi)\equiv (\pm2\delta(E),0)$. With
$\delta(E)$ being a decreasing function of energy, one naively
expects LDOS modulation wavevectors to decrease at a higher bias
as well, in qualitative agreement with the
experiment\cite{hoffman2}. Below, we examine through explicit
calculation whether this naive argument will hold.

The treatment can proceed by positing the effective action
$\cL=\cL_M +\cL_{qp}+\cL_{qp-M}$ consisting respectively of the
magnetic and quasiparticle sectors and their coupling. For the
magnetic part\cite{pvs}, \ba \cL_M [\Phi,\Phi^*]&=&{1\over2}\int
d^3 x d^3 y\Phi^* (x) \chi^{-1}(x-y)\Phi(y) \nn &-&{1\over2}\!\int
d\tau [\xi^* \Phi^2 (\v r_i,\tau)+\xi\Phi^{*2} (\v r_i,\tau)]. \ea
Space-time coordinates are denoted by $x=(\v r,\tau)$ and $y=(\v
r',\tau')$. The last term is the pinning potential, located around
$\v r_i$, for the spin density wave(SDW). Due to the pinning term
the propagator $\langle \Phi^* (\v r, i\omega)\Phi(\v r',
i\omega)\rangle\equiv \chi(\v r,\v r',i\omega)$ is no longer
translationally invariant.

The quasiparticle coupling to the fluctuating SDW is modelled
by\cite{pvs} \be \cL_{qp-M} ={1\over2}g S(\v r\tau)\cdot[\Phi(\v r
\tau)+\Phi^* (\v r \tau)].\ee Through this term the quasiparticle
propagator is influenced by the collective mode, and exhibits
translational symmetry breaking. The perturbative treatment of the
effective interaction is straightforward, with the one-loop result
for the change in Green's function $\delta\cG$\cite{comment}, \ba
&&\delta\cG (\v r,\v r',i\omega)=\sum_{\v r_1, \v r_2,i\nu}g^2
(i\nu) \chi(\v r_1\!-\!\v r_i, i\nu)T(i\nu)\chi(\v r_i \!-\! \v
r_2,i\nu) \nn &&~~~~~~\times G(\v r\!-\!\v r_1, i\omega)G(\v r_1
\!-\!\v r_2,i\omega\!-\!i\nu)G(\v r_2\!-\!\v
r',i\omega),\label{dG_M}\ea where $T(i\nu)=\chi(\v r = 0,i\nu)$.
Frequency dependence in the coupling $g(i\nu)$ is introduced for
generality. In arriving at this expression we have only kept those
terms that break the translational symmetry and ignored the
space-independent self-energy correction. After the Wick rotation,
$i\omega\rightarrow\omega+i\delta$, the spatially varying LDOS is
$N(\v r,\omega)=(1/\pi){\rm Im}\delta\cG(\v r,\v
r,\omega+i\delta)$.

Normally the summation $\sum_{i\nu}$ is carried out analytically,
and the resulting expression is evaluated by numerical means after
the Wick rotation $i\omega\rightarrow\omega+i\delta$. In our case
the renormalized susceptibility $\chi(\v q,i\nu)$ does not possess
a simple Lehmann expansion, which prevents the frequency summation
from being carried out. Instead we resort to the time-ordered
Green's function method (as opposed to the Matsubara Green's
function method) which does not require the Wick rotation. At zero
temperature LDOS can be obtained from the time-ordered Green's
function by ($T=$time-ordered)\be N(\v
r,\omega)=(1/\pi)\sgn(\omega) {\rm Im}\delta\cG_T (\v r,\v
r,\omega).\ee Perturbation theory proceeds in entirely analogous
manner, and we only quote the final result for the Green's
function correction.\ba &&\delta\cG_T (\v r,\v r',\omega)=
-i\sum_{\v r_1, \v r_2}\int_{\nu}g^2 (\nu)\chi_T(\v r_1\!
\!-\!\!\v r_i, \nu)T(\nu)\chi_T(\v r_i \!\!-\!\! \v r_2,\nu) \nn
&& ~~~~~\times G_T(\v r\!-\!\v r_1, \omega)G_T(\v r_1 \!\!-\!\!\v
r_2,\omega\!-\!\nu)G_T (\v r_2\!\!-\!\!\v
r',\omega).\label{dG_T}\ea The time-ordered susceptibility
$\chi_T$ and Green's function $G_T$ are obtained from their
Matsubara counterparts by $i\omega\rightarrow
\omega+i\sgn(\omega)$.
\\

{\it Results}: We evaluate Eq. (\ref{dG_T}) numerically on a large
lattice of dimension [201$\times$201] using self-consistently
determined parameters $\Delta_{max} \equiv 2|\langle
\epsilon_{\alpha\beta}f_{i\alpha}f_{j\beta}\rangle|=0.48J$. The
impurity site is located at the center, $\v r_i = (101,101)$. We
use $J_{eff}=0.34J$, and the broadening factor $\delta=0.01J$ in
evaluating $\chi(\v r,\nu)$ and $G(\v r,\omega)$. The
$\nu$-integration is replaced with a discrete sum, $\int d\nu
f(\nu)\rightarrow \Delta\nu\sum_s f(s\Delta\nu)$, where $s$ is an
integer, $-s_{max}\le s \le s_{max}$. The following results are
taken with $\Delta\nu=0.04J$, and $s_{max}=14$ implying that
virtual processes with the off-shell energy of up to $0.56J$ are
taken into account in the self-energy evaluation. The commensurate
response at $\v q =(\pi,\pi)$ in the susceptibility occurs at
$\nu_{res}\approx 0.32J$ for the present model, while at energies
above and below $\nu_{res}$, the resonance occurs away from
$(\pi,\pi)$\cite{bl}. Our self-energy energy calculation therefore
includes effects from both above and below the resonance energy.
We have confirmed on a smaller lattice that using a finer scale
$\Delta\nu=0.02J$ does not change the final result qualitatively.
The on-site Green's function $\delta\cG_T(\v r,\v r,\omega)$ is
evaluated for a subset of the lattice, typically for $\v r$ lying
in [101$\times$101] about the impurity center, for $|\omega|$ up
to $0.32J$. The real-space LDOS thus obtained is then
Fourier-transformed according to $N(\v q,\omega)=\left|\sum_{\v
r}e^{-i\v q\cdot \v r}N(\v r,\omega)\right|$.

\begin{figure}
\centering \hskip -1.0cm
\epsfxsize=9cm\epsfysize=8cm\caption{Fourier-transformed LDOS
$N(\v q,\omega)$ for the $\v q$-range $[-\pi,\pi]\times[\pi,\pi]$
and for $-0.32J \le\omega \le 0.32J$. Center of each rectangle
corresponds to $\v q = (0,0)$. The intensity scheme is with
respect to each figure, and should not be compared between
different plots. ({\bf Please refer to attached figure
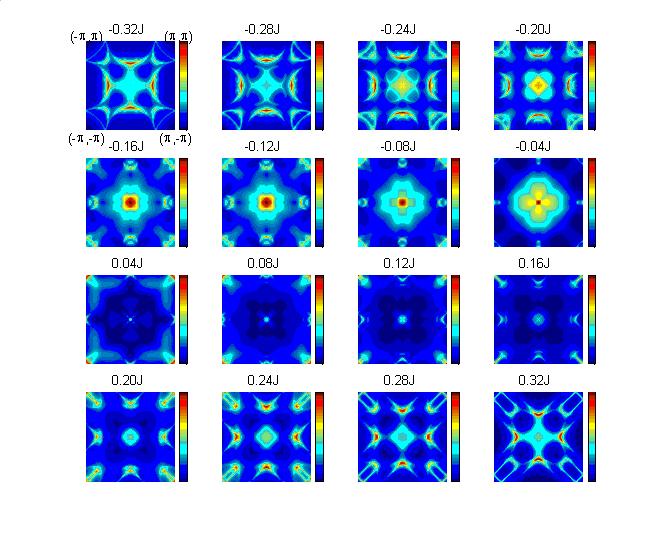})} \label{N_kw}
\end{figure}

\begin{figure}
\centering \hskip 0.0cm\epsfxsize=8cm\epsfysize=4.5cm
\epsfbox{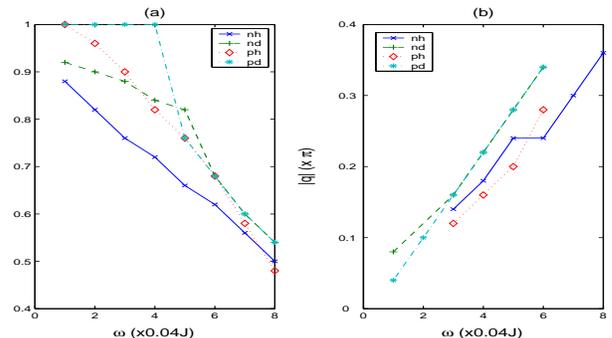}\caption{Energy dependence of $q_d (\omega)$
and $q_h (\omega)$ (defined in the text) obtained from $N (\v
q,\omega)$ (Figure (a)) and from $N_{red}(\v q,\omega)$ (figure
(b)). In each figure, four data sets are shown for $q_d
(\omega<0)$ (nd), $q_h (\omega<0)$ (nh), $q_d (\omega>0)$(pd), and
$q_h (\omega>0)$(ph) against $|\omega|$. Note that all the $q$
values decrease with energy $|\omega|$ in (a), and increase in
(b).} \label{k-vs-E}
\end{figure}

Plots in Fig. \ref{N_kw} show $N(\v q,\omega)$ calculated with the
coupling constant $g^2(\nu)=$const. Apart from the $\v q=0$ peak,
we find some prominent peaks along both diagonal (meaning
$(\pm\pi,\pm\pi)$) and horizontal (meaning $(\pm\pi,0)$ and
$(0,\pm\pi)$) directions. The peaks are located near the ends of
the Brillouin zone at a small bias, and move toward $\v q=0$ as
the bias energy is increased. A horizontal(diagonal) scan of $N(\v
q,\omega)$ allows the identification of wavevectors $\v q_h
(\omega)$ ($\v q_d (\omega)$) for which $N(\v q,\omega)$ is a
local maximum. The values of $q_d (\omega)\equiv |\v q_d (\omega)
|/\sqrt{2}$ and $q_h (\omega) \equiv |\v q_h (\omega) |$ are
plotted in Fig. \ref{k-vs-E}(a). Clearly both $q_d (\omega)$ and
$q_h (\omega)$ are decreasing functions of energy. The detailed
shape of $N(\v q,\omega)$ however depends somewhat on whether $V$
is negative (electrons tunneling out) or positive (electrons
tunneling in).

\begin{figure}
\centering \hskip -0.0cm \epsfxsize=8cm\epsfysize=8cm
\caption{LDOS $N_{red}(\v q,\omega)$ calculated with the reduced
magnetic susceptibility $\chi_{red}$. Energy and momentum ranges
are the same as in Fig. \ref{N_kw}. ({\bf Please refer to attached
figure 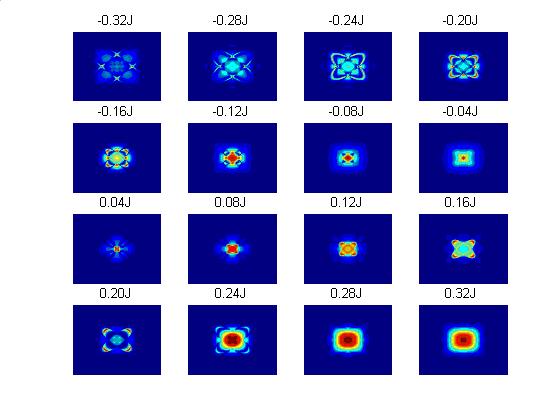})}
\label{N_kw_red}
\end{figure}

As shown in Fig. \ref{chi-k} the magnetic absorption peaks are
most pronounced around $(\pi,\pi)$. Note however that the Green's
function shown in Eqs. (\ref{dG_M}) and (\ref{dG_T}) are a product
of several complicated functions, and the magnetic absorption
spectrum alone does not a priori characterize the LDOS. In this
regard we can loosely classify the structures in $\chi''(\v
q,\nu)$ as those near $(\pi,0),(0,\pi)$, and those near
$(\pi,\pi)$. In the result shown in Fig. \ref{N_kw} contributions
from these two regions will undoubtedly be mixed. Instead, if we
truncate the magnetic spectrum to be within a certain vicinity of
$(\pi,\pi)$, the quasiparticle Green's function will also be
influenced by the magnetic excitations at these wavevectors, and
none from around $(\pi,0),(0,\pi)$. This is achieved by using the
reduced susceptibility $\chi_{red}(\v q,\nu)=\chi(\v q,\nu)$ for
$-\pi/4<q_x ,q_y <\pi/4$, $\chi_{red}(\v q,\nu)=0$ otherwise. Its
Fourier transform $\chi_{red} (\v r,\nu)$ is used in the
evaluation of the new LDOS, $N_{red}(\v r,\omega)$ and $N_{red}
(\v q,\omega)$, shown in Fig. \ref{N_kw_red}. In this calculation
we use the cutoff $g^2(\nu)=g^2(0)\exp [-0.1 (\nu/\Delta\nu)^2]$.
For a less sharp cutoff such as $g^2 (\nu)=g^2 (0)$ or
$g^2(\nu)=g^2(0)\exp[-0.1 |\nu/\Delta\nu|]$, we find no other
distinguishable structure in $N_{red} (\v q,\omega)$ except a
broad hump centered at $\v q =0$.

There is very little overlap between the two figures \ref{N_kw}
and \ref{N_kw_red}, as also reflected in the $q_d (\omega)$ and
$q_h (\omega)$ values obtained respectively from each figure.
Figure \ref{k-vs-E}(b) shows the peak intensity position of
$N_{red}(\v q,\omega)$. Essentially all $q_d (\omega), q_h
(\omega)$ are greater than $\pi/2$ in the $N(\v q,\omega)$ while
in the reduced LDOS, all $q_d (\omega)$, and $q_h (\omega)$ are
less than $\pi/2$. Furthermore, $q_d(\omega)$ and $q_h (\omega)$
in Fig. \ref{k-vs-E}(b) are increasing functions of energy, as
opposed to Fig. \ref{k-vs-E}(a) where they are decreasing. In both
plots, $q_d (\omega)$ and $q_h (\omega)$ show the same dependence
on energy, being both increasing or both decreasing functions.  As
previously mentioned a standard Ginzburg-Landau argument about the
coupling between SDW and charge-density-wave (CDW) \cite{GL}
implies that an ordering tendency occurring at
$(\pi\pm\delta,\pi\pm\delta)$ for the SDW will lead to CDW
modulations at $(\pm 2\delta,\pm 2\delta)$. Naively, since
$\delta$ decreases with energy, one expects $q_h (\omega)$ in Fig.
\ref{k-vs-E}(b) to be a decreasing function of $\omega$, which is
contradicted by our calculation. In obtaining Fig. \ref{N_kw_red}
we (1) truncated the magnetic spectrum within a small momentum
window around $(\pi,\pi)$ and (2) modified the coupling
$g^2(\nu)$. With this manipulation we were able to bring out
features in the LDOS which were ``hidden" in the calculation that
led to Fig. \ref{N_kw}. Since the coupling constant $g$ must in
reality be a function of both momentum and energy, one cannot be
too certain a priori which of the features found in Figs.
\ref{N_kw} and \ref{N_kw_red} are more readily observable in an
experiment.

{\it Discussion}: We now come to comparison of our result with the
experimental data. First, it has been found in Ref.
\cite{hoffman2} that while $q_d (\omega)$ increases, $q_h
(\omega)$ decreases at a larger bias $|\omega|$. This is in
contrast to our calculation where both $q_d (\omega)$ and $q_d
(\omega)$ behave analogously with respect to energy. Secondly,
experimentally measured $k_h (\omega)$ values are found to lie
below $\pi/2$ for a wide range of energies and hole doping, which
may be consistent with Figure \ref{k-vs-E}(b), but in this case
the energy dependence is opposite to our calculation. In Fig.
\ref{k-vs-E}(a), $q_d (\omega)$ and $q_h (\omega)$ values are
rather high which, translated into real space, implies LDOS
modulations on the length of $2-4$ lattice spacings. While such
magnetic-fluctuation-induced LDOS variations are not forbidden and
may very well be observable in the future, we must conclude that
the LDOS modulation based on the $t-t'-J$ model of the magnetic
excitation is inconsistent with the currently known
experiment\cite{hoffman2}.

It is possible that the renormalized $t-t'-J$ model does not after
all capture the quasiparticle band structure of the BSCCO and that
in another model, one indeed finds the energy dependence of $q_d$
and $q_h$ consistent with the experiment within the magnetic
fluctuation scenario. For example, in WL\cite{wl}, the model
adopted is the phenomenological tight-binding model originally
proposed by Norman\cite{norman}.

On the theoretical side, one should be careful that the naive
Ginzburg-Landau argument for SDW-CDW coupling may in some cases
lead to predictions which are inconsistent with a full many-body
calculation such as this one. Although the patterns in $N(\v
q,\omega)$ are ultimately due to the underlying magnetic
fluctuation, the reason for the particular energy dependence of
the modulation period found in Figs. \ref{N_kw} and \ref{N_kw_red}
remains unclear. Finally an entirely different mechanism such as
WL's are not incompatible with the present model, and may well
simultaneously lead to observable effects in a given system.

We thank Seamus Davis, Dung-Hai Lee, and Qiang-Hua Wang for
reading the manuscript and for critical comments. We acknowledge
the financial support from the Faculty Fund of Konkuk University.
This work is also supported in part by the Korean Science and
Engineering Foundation (KOSEF) through the Center for Strongly
Correlated Materials Research (CSCMR).

\widetext


\begin{thebibliography}{99}
\bibitem{comment} Actual result involves a multiplicative factor
of order unity, which we absorb into the definition of the
coupling $g$.
\end{thebibliography}
\end{document}